\documentclass[mathleft
]{an}
\usepackage{graphicx}
\usepackage{times}
\usepackage{amssymb}
\usepackage{amsmath}
\begin{document}

% The following seven commands are intended for editorial usage and should be ignored by
% the author(s).
\Pagespan{1}{}% Document's page range. 
% If second parameter is left empty, the last page is computed automatically.
\Yearpublication{2016}%
\Yearsubmission{2016}%
\Month{11}%   
\Volume{xxx}%  
\Issue{xx}% 
% \DOI{This.is/not.aDOI}% 

\title{CHEERS: Future perspectives for abundance measurements in clusters with \textit{XMM-Newton}}

\author{J. de Plaa\inst{1}\fnmsep\thanks{Corresponding author:
  \email{j.de.plaa@sron.nl}\newline}
%Example 
%for footnote, note the usage of the \texttt{fnmsep}
%command as separator between institute number and footnote mark} 
\and  F. Mernier\inst{1,2}
}
\titlerunning{}
\authorrunning{J. de Plaa \& F. Mernier}
\institute{
SRON Netherlands Institute for Space Research, Sorbonnelaan 2, 3584 CA Utrecht, The Netherlands
\and 
Leiden Observatory, Leiden University, P.O. Box 9513, 2300 RA Leiden, The Netherlands
}

\received{September 23, 2016}
\accepted{November 4, 2016}
\publonline{later}


\abstract{%
The CHEERS (CHEmical Enrichment RGS Sample) observations of clusters of galaxies with \textit{XMM-Newton} have shown to be valuable to constrain the chemical evolution of the universe. The soft X-ray spectrum contains lines of the most abundant metals from N to Ni, which provide relatively accurate abundances that can be compared to supernova enrichment models. The accuracy of the abundances is currently limited by systematic uncertainties introduced by the available instruments and uncertainties in the modeling of the spectra, which are of the order of 20$-$30\%. We discuss the possible gain of extending the current samples at low and high redshift. We conclude that expanding the samples would be expensive in terms of exposure time, but will not yield significantly improved results, because the current samples already reach the systematic limits. New instrumentation, like \textit{Astro-H2} and \textit{ATHENA}, and improvements to the atomic databases are needed to make significant advances in this field.  
 }

\maketitle

\section{Introduction}

In the first few minutes of the universe, 'Big-Bang' nucleosynthesis produced mainly helium and traces of lithium and beryllium by nuclear fusion of protons. As the universe expanded, the primordial gas cooled down and the fusion reactions ceased. About 500 Myr after the 'Big Bang', the first generation of metal-poor stars formed from the cooled primordial gas. This first generation of stars (also named Population III stars) is thought to consist of massive stars (M $\gtrsim$ 40 M$_{\sun}$; Bromm et al. 2004)\nocite{bromm2004}. Although these stars were the first to produce low-mass metals like carbon and oxygen, they do not contribute a lot to the metalicity (10$^{-4}$ Z$_{\mathrm{\sun}}$, Matteucci \& Calura 2005)\nocite{matteucci2005}.  

Only when the star formation rate reached its peak around a redshift of $z=2-4$ (e.g. Madau \& Dickinson, 2014)\nocite{madau2014}, the bulk of the metals were produced. Galactic winds ejected a substantial fraction of the produced metals into the surrounding medium. In the most massive objects, clusters of galaxies, the gas ejected by the galactic winds mixes with in-falling intergalactic gas from the cluster surroundings. Due to the gravitational energy release of the in-falling gas and the shocks produced by mergers between the cluster and galaxies or sub-clusters, the intra-cluster medium (ICM) is heated to temperatures $>$10$^6$ K. Around $z=2$ the ICM is hot enough to emit X-rays. The presence of hot gas in the cluster causes in-falling galaxies to loose their cool gas component by ram-pressure stripping and suppresses star formation in the cluster galaxies. Therefore, most of the member galaxies are left with an old stellar population.

A large fraction of the metals has been produced in the peak of the star formation rate by core-collapse and type Ia supernovae (SNcc and SNIa). From $z \sim 2$, the formation of the hot ICM is thought to have consequences for the chemical enrichment of the gas. Since it has become difficult to form new stars, the progenitors of core-collapse supernovae, massive stars (M $\gtrsim$ 8 M$_{\sun}$), quickly become rare. Type Ia supernovae, on the other hand, need one or more white-dwarf progenitors in a binary, which continue to form in the galaxies because they originate from long-lived low and intermediate mass stars. After $z \sim 2$, the core of the cluster is therefore mainly enriched with type Ia products, which provides a slightly different chemical enrichment history compared to our own galaxy, where star formation and core-collapse supernova enrichment continued.         

The bulk of the metals produced by supernovae show spectral lines in the soft X-ray spectrum emitted by the hot ICM. The X-ray spectra contain, among others, lines of oxygen, neon, and magnesium, which are typical SNcc products (e.g. Woosley \& Weaver 1995; Nomoto et al. 2006)\nocite{woosley1995,nomoto2006}, and argon, calcium, iron, and nickel lines, which are typical products of SNIa (e.g. Iwamoto et al. 1999; Bravo \& Mart\'inez-Pinedo 2012). \nocite{iwamoto1999,bravo2012} The soft X-ray band is therefore a very useful energy band to do chemical enrichment studies. Since the hot ICM plasma is in (or close to) collisional ionization equilibrium, the X-ray spectrum and the metal abundances are relatively easy to model, which potentially enables accurate measurements of the abundances in clusters of galaxies. A more extensive review of measuring and interpreting abundances in the ICM can be found in e.g. de Plaa (2013), B\"ohringer \& Werner (2010), and Werner et al. (2008)\nocite{deplaa2013,boehringer2010,werner2008}.  

\textit{XMM-Newton} has so far provided valuable contributions to the field of clusters of galaxies. The question is whether \textit{XMM-Newton} can continue to play a role in cluster abundance studies in the coming years. We summarize the results from the CHEmical Enrichment RGS Sample (CHEERS) of deep \textit{XMM-Newton} observations of clusters of galaxies. We focus on the current limitations of measuring abundances. We discuss what is needed to significantly improve abundance measurements in clusters of galaxies in X-rays and how \textit{XMM-Newton} and future missions like \textit{Astro-H2} and \textit{ATHENA} can contribute to this field.   

\section{The CHEERS project}

The CHEERS project was conceived with the acceptance of a Very Large Program (VLP) with \textit{XMM-Newton} in AO-12. The aim of the proposal was to obtain a 'complete' sample of clusters of galaxies that are optimal for observation with the Reflection Grating Spectrometer (RGS) aboard \textit{XMM-Newton}. Together with archival observations, the CHEERS collaboration compiled a sample of 44 clusters with deep RGS exposures (de Plaa et al., 2016, subm.). \nocite{deplaa2016} Due to the design of the RGS instrument, the cluster selection is biased favoring centrally peaked clusters. Since RGS does not contain a slit, the spatial extent of the source translates into an effective broadening of the measured spectral lines. A peaked surface brightness profile therefore provides the sharpest lines, while a flat surface brightness profile causes the lines to be broadened extensively and makes the lines harder to resolve. Clusters were selected if an observation with RGS would allow a 5$\sigma$ detection of the O VIII Ly$\alpha$ line, and if $z \le 0.1$. This resulted in the selected set of observations with a total effective exposure time of $\sim$4.5 Ms.

\subsection{Chemical enrichment}   
\label{sec:chem_enrichment}

To measure abundances in the ICM, both RGS and EPIC can be used. The RGS instrument is better suited to measure oxygen and neon, because EPIC lacks the spectral resolution to resolve the neon line in the Fe-L complex and the oxygen line is uncertain in EPIC due to the nearby oxygen edge. The EPIC instrument is well suited to measure Mg, Si, S, Ar, Ca and Ni abundances, because it covers a larger energy window (up to $\sim$ 10 keV) and can thus access the K-shell emission lines of these elements. The Fe abundance determination using EPIC is more accurate, as it can be constrained by both the Fe-L and the Fe-K complexes. De Plaa et al. (2016, subm.)\nocite{deplaa2016} present the measurements of the O/Fe ratio using RGS and Mernier et al. (2016a)\nocite{mernier2016a} discuss the EPIC abundance measurements. Both papers also discuss the statistical and systematic uncertainties in these abundances. Especially, the uncertainties in the spectral modeling appear to be important (see also Section~\ref{sec:atomiclim}).

The best abundance sets of RGS and EPIC are compared to more recent supernova models in Mernier et al. (2016b)\nocite{mernier2016b}. They confirm a significant underestimate of the Ca/Fe and Ni/Fe ratios predicted by the models. The predicted Ca/Fe ratio can be reconciled with the observations when assuming either a SNIa model that reproduces the spectral features of the Tycho supernova remnant (Badenes et al. 2006\nocite{badenes2006}; see also de Plaa et al. 2007\nocite{deplaa2007}), or a significant contribution from calcium-rich gap transients (a recent subclass of supernovae that produces relatively more Ca and explodes in the outskirts of their host galaxies; e.g. Waldman et al. 2011\nocite{waldman2011}). On the other hand, the Ni/Fe ratio matches the observations when assuming two explosion channels for SNIa (deflagration and delayed-detonation). 

\subsection{Additional science}

Apart from chemical enrichment studies, the deep RGS data of the CHEERS sample can also be used to study other aspects of clusters. Pinto et al. (2015) \nocite{pinto2015} study the turbulence in the ICM. By carefully modeling the line broadening in the RGS instrument, they were able to find upper limits for turbulent motions in the gas. In principle, RGS also allows to measure turbulence through resonant scattering of Fe XVII lines. Using this effect, Ahoranta et al. (2016) \nocite{ahoranta2016} find evidence for asymmetry in the turbulent gas properties of NGC 4636. In both studies, it is also essential to consider systematic uncertainties to estimate the accuracy of the results.    

Pinto et al. (2014)\nocite{pinto2014} discovered the presence of a low-temperature component in the cores of clusters using CHEERS RGS data. Before, the lowest temperatures in the ICM were identified using the detection of Fe XVII lines, which indicate a temperature range of $\sim$0.5-0.8 keV. In a few clusters and groups, line emission from O VII was detected indicating a temperature of $\sim$0.2 keV. The line ratios in the O VII triplet appear to be affected by resonant scattering, which suggests that the turbulence in the gas is low.   

\section{Systematic uncertainties \& limitations}

The clusters in the CHEERS sample are observed with average exposure times of $\sim 100$ ks. This results in data sets with excellent statistical quality. In this regime, the uncertainties are not dominated by statistical noise anymore, but by systematic uncertainties in various components of the analysis. This does not necessarily mean that performing long exposures is pointless, because the low statistical noise allows a more detailed study of systematic effects. In some cases, it is possible to identify and understand the source of a systematic error. If a systematic uncertainty can be eliminated this way, the results can improve substantially.

Systematic errors can be introduced in various ways. There can be errors in the initial assumptions, (hidden) uncertainties in the used models, errors in the data reduction, and errors in the calibration of the instrument. In this section, we discuss the main systematic uncertainties and their impact on abundance measurements and their interpretation. These systematic uncertainties are effectively limitations that need to be solved in order to advance this field.

\subsection{Limitations of the RGS spectrometer}

X-ray Grating spectrometers, like RGS, are optimized for point-source spectroscopy. However, RGS is one of the few X-ray gratings that allow limited, but meaningful, analysis of extended sources. One of the requirements is that the surface brightness of the cluster needs to be centrally peaked. Due to the slit-less design of RGS, the line-spread function mainly depends on the surface brightness of the source along the dispersion axis of the instrument. The line-spread function is even different for each ion, because the surface brightness of the line depends on the abundance and temperature gradient in the plasma. This makes it hard to model the spectrum. In the spectral fits, the model is broadened by one or two broadening kernels derived from an EPIC MOS image of the source. According to de Plaa et al. (2016)\nocite{deplaa2016}, the systematic error introduced by this approximation of the line broadening is about 10\% in the O/Fe abundance. 
 
Because of the spatial line broadening, the spectral resolution of RGS is effectively degraded, which makes it more difficult to resolve and detect weak lines. Checks with simulated spectra have shown that the abundance is mainly constrained by the flux in the core of the bright lines, which limits the systematic uncertainty to a (for now) acceptable level of 10\%. However, using RGS data it is not easy to detect weak lines, like Ni, Na, and Al lines which are blended with the Fe-L complex. A non-dispersive spectrometer would be needed to resolve these lines. 

Although RGS has the highest effective area of the grating spectrometers in orbit, it is still limited in effective area compared to the CCD imaging spectrometers. Relatively long exposure times of $\sim$100 ks are needed to obtain good quality spectra of local clusters. In addition, the RGS bandwidth of 0.1$-$2.5 keV is quite limited and becomes ineffective for clusters with central temperatures above $\sim$5 keV. Also the determination of $N_\mathrm{H}$ is uncertain in cluster spectra, because the absorption line features are also spatially broadened and the bandwidth does not cover the high-energy range that is not affected by absorption. This leads to uncertainties in the continuum level determination, which in turn affects the absolute abundance measurements.

\subsection{Limitations of the EPIC spectrometer}
\label{sec:EPIC}

\begin{figure}[!t]
        \centering
                \includegraphics[width=0.487\textwidth]{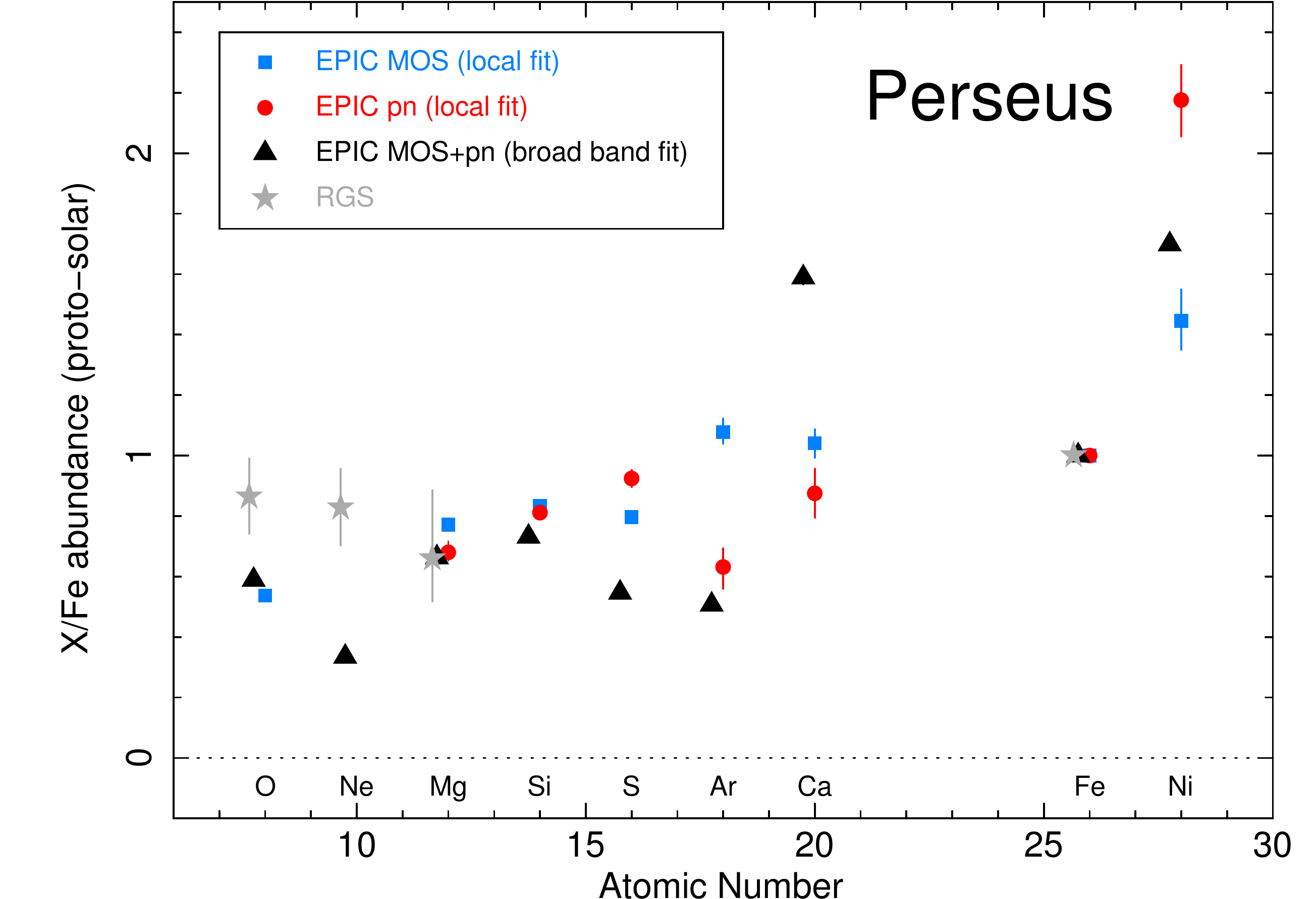}

        \caption{Abundance ratios (X/Fe) measured with RGS and EPIC instruments in the Perseus cluster.}
\label{fig:EPIC_Perseus}
\end{figure}

Due to the broad band of EPIC, broad-band modeling sometimes over- or underestimates the local continuum due to calibration or model uncertainties, which causes bias in the abundance determination. Mernier et al. (2016a)\nocite{mernier2016a} therefore fit the abundance locally in a narrow energy band around the line. Fig.~\ref{fig:EPIC_Perseus} shows the abundances measured in the Perseus cluster ($<$0.2 $R_{500}$). The abundances from the broad-band fit (black triangles, MOS+pn) show substantial differences with the local narrow band fits (blue squares and red circles, for independent MOS and pn fits, respectively). The plot shows that ignoring broad-band effects severely biases the S/Fe and the Ca/Fe measurements in Perseus (31\% and 53\%, respectively).

Moreover, the high statistics of the Perseus observations reveal significant discrepancies between MOS and pn measurements, even after correction from broad-band effects. These MOS-pn discrepancies are particularly striking for Ar/Fe ($\sim$41\%) and Ni/Fe ($\sim$50\%), and illustrate the current limitations of the EPIC spectrometer. This is probably related to uncertainties in the effective area calibration (see also Schellenberger et al. 2015)\nocite{schellenberger2015} and/or to difficulties to correctly model the non-X-ray background.

As mentioned in Section~\ref{sec:chem_enrichment}, the moderate spectral resolution of EPIC does not allow to derive O and Ne abundances with a good accuracy (see the EPIC vs. RGS discrepancies in Fig.~\ref{fig:EPIC_Perseus}). Furthermore, the K-shell transitions of Mg ($\sim 1.5$ keV) are near the Al K$\alpha$ fluorescent instrumental line present in both MOS and pn instruments ($\sim 1.49$ keV). The determination of Mg is reliable in cluster cores, where the background is low, but it becomes challenging toward the outskirts, which show a larger background contribution.

\subsection{Limitations of the spectral modeling}
\label{sec:atomiclim}

The most hidden systematic uncertainties are the differences in spectral emission models. Since it is difficult to get robust absolute atomic data from laboratory measurements, the current CIE models from SPEX and APEC depend mostly on atomic data calculated from atomic models. These also contain uncertainties that are not easy to quantify. In order to get an impression of the magnitude of the systematic uncertainty in RGS, de Plaa et al. (2016)\nocite{deplaa2016} estimate the differences in the O/Fe abundance ratio between SPEX and APEC. If the old SPEX CIE model (SPEXACT v2) is compared to APEC, the differences in the O/Fe abundance can be as large as 50\%. Using the updated CIE model of SPEX (SPECACT v3) the differences between APEC and SPEX are reduced to $\sim$20\% at maximum. This suggests that at least the O/Fe abundance ratio from theory is not known better than 20\%.

Next to the internal model uncertainties, there are also uncertainties related to the choice of models applied to the source. Is there multi-temperature structure, a contribution from AGN, is the N$_H$ column well determined, and is there dust along the line of sight? The uncertainty in multi-temperature modeling on the O/Fe ratio is approximately 20\% at maximum (de Plaa et al., 2016)\nocite{deplaa2016}. On top of that, the best-fit N$_H$ may be different from the N$_H$ found by Radio measurements (Schellenberger et al., 2015; Willingale et al., 2013)\nocite{schellenberger2015,willingale2013}, which causes a shift in the O/Fe ratio which roughly scales with the difference in absolute N$_H$. Roughly, the relative change in O/Fe is about 20\% per 3 10$^{20}$ cm$^{-2}$ difference in N$_H$. This means the O/Fe is most sensitive in objects with a large N$_H$ value.

\section{Future prospects}

\subsection{\textit{XMM-Newton}}

The amount of clusters that can be successfully probed with RGS is limited to bright, centrally peaked and cool clusters, which makes it difficult to significantly enlarge the sample with observations of other clusters. One could consider extending the sample further with deep EPIC observations, but with $\sim 4.5$ Ms of cleaned data, we would require more than 20 Ms to improve the statistics by a factor of two (taking into account the loss of exposure due to soft protons). However, the CHEERS sample has shown that the systematic uncertainties already dominate over the statistical uncertainties. The instrumental limits are reached in the current sample and adding more nearby clusters will not significantly improve the current achievements. Improvements can only be expected from the improvement of atomic data, spectral models, and new high-resolution X-ray spectrometers.

\subsection{Probing recent ICM enrichment with redshift}

Another option to study chemical evolution is to measure abundances as a function of redshift. This option allows to trace the evolution of the enrichment with time, but due to the large distances of the objects, the analysis is limited mainly to the evolution of Fe. A first \textit{ASCA} study did not show evidence of evolution in the ICM metalicity up to $z \sim 0.4$ (Mushotzky \& Loewenstein 1997)\nocite{mushotzky1997}. With \textit{XMM-Newton} and \textit{Chandra} clusters were discovered up to $z \sim 1.3$ and samples of clusters showed a hints of an increasing abundance with time (Balestra et al. 2007; Maughan et al. 2008; Anderson et al. 2009)\nocite{balestra2007,maughan2008,anderson2009}. However, this trend was not confirmed (Tozzi et al. 2003; Baldi et al. 2012)\nocite{tozzi2003,baldi2012}, essentially due to intrinsic scatter in the measurements. Moreover, the statistical errors on the metalicities of higher-$z$ clusters are often large ($\gtrsim 30 \%$), and the question of whether deeper EPIC exposures would significantly improve the current picture clearly arises.

\begin{figure*}[!t]
        \centering
                \includegraphics[width=0.487\textwidth,trim={11 15 0 0},clip]{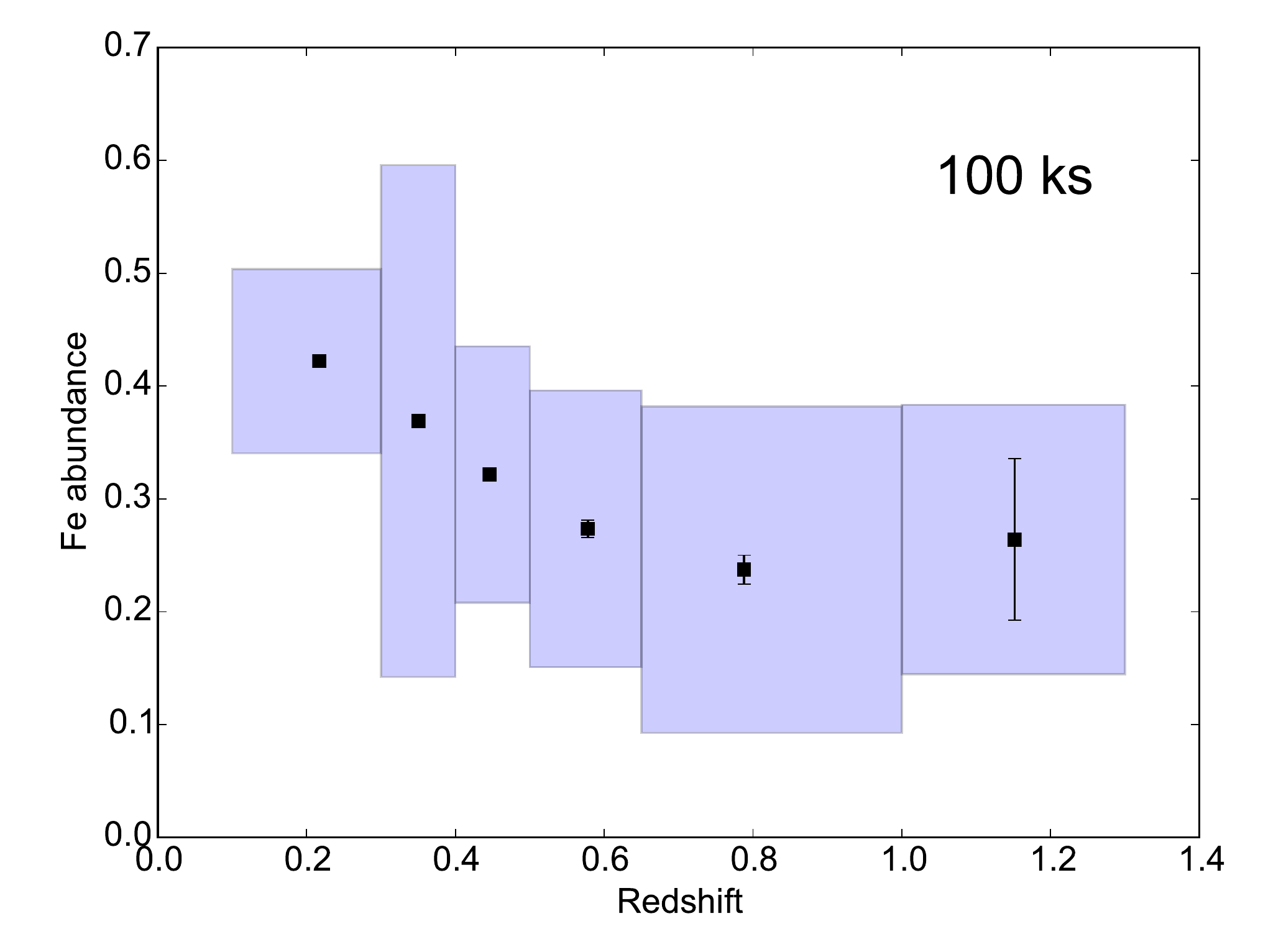}
                \includegraphics[width=0.487\textwidth]{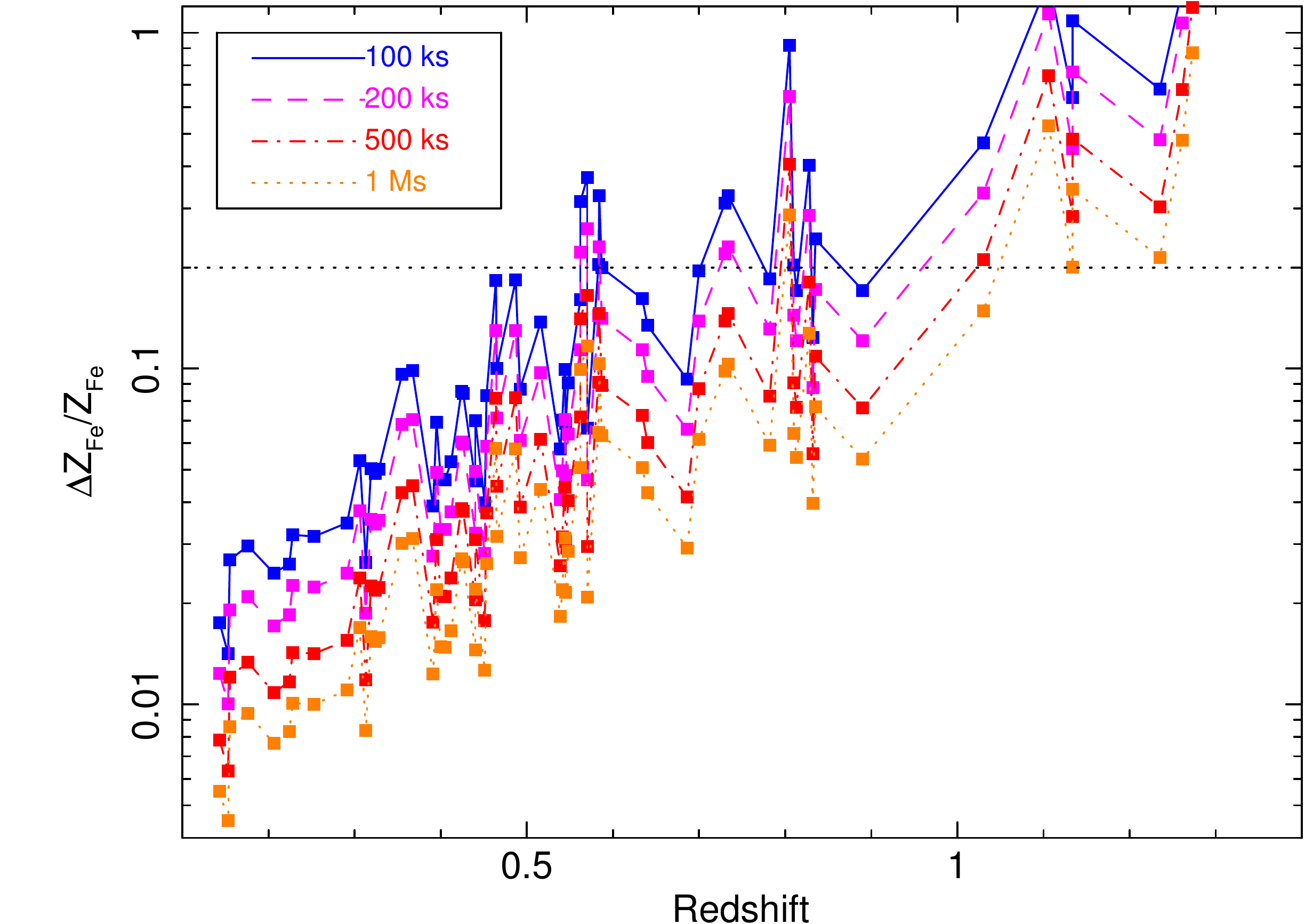}

        \caption{EPIC simulations of Fe abundance determination of distant clusters, based on the sample of Balestra et al. (2007)\nocite{balestra2007}. \textit{Left:} Determination of the average evolution of Fe abundance with redshift, assuming 100 ks of net exposure for each cluster.  The blue shaded areas show the rms dispersion around the weighted mean. \textit{Right:} Relative statistical errors ($\Delta Z_\text{Fe} / Z_\text{Fe}$) on the metalicity for each cluster from the sample. The simulations assume successively 100 ks, 200 ks, 500 ks and 1 Ms of net exposure. The horizontal dotted line indicates the $\Delta Z_\text{Fe} / Z_\text{Fe} = 20 \%$ threshold, corresponding roughly to the scatter of Fe abundance in clusters (Mernier et al. 2016a)\nocite{mernier2016a}.}
\label{fig:simu_Balestra}
\end{figure*}

To estimate whether new high-redshift cluster data is useful for chemical evolution studies, we consider extending the sample of Balestra et al. (2007)\nocite{balestra2007}. This sample consists of 65 clusters between $0.1 \lesssim z \lesssim 1.3$ that are observed with \textit{Chandra}. For each cluster, we use the best-fit values for the temperature and abundance from Balestra et al. (2007; see their Table 3)\nocite{balestra2007} as input for a simulation with the SPEX fitting package (Kaastra et al. 1996)\nocite{kaastra1996}. For each cluster, we simulate MOS and pn spectra with exposure times of 100 ks per cluster. The derived Fe abundances, averaged over six redshift bins, are shown in the left panel of Fig. \ref{fig:simu_Balestra} ("best-fit" sample, blue squares). The blue shaded areas represent the root mean square (rms) dispersion around the weighted mean in each bin that was found in the original sample. Although the combined statistical uncertainties in each bin are small, the large rms dispersion prevents us from claiming any significant decrease of metalicity with $z$.

It is important to note that we have included X-ray and non-X-ray background components in our simulations. Especially when the net flux of the ICM emission is weak, the non-X-ray background may dominate at high energies ($\gtrsim 2$ keV), making the Fe abundance determination via K-shell transitions difficult. This applies for the most distant clusters, as illustrated in the right panel of Fig. \ref{fig:simu_Balestra}, where we show the relative statistical uncertainty in metalicity as a function of redshift and exposure time. In any $z \ge 1$ cluster, even 1 Ms of EPIC net exposure would not be sufficient to constrain the metalicity with an accuracy of $20 \%$ or less.

How many clusters should we observe to be able to significantly distinguish abundance evolution as found by Balestra et al. (2007)\nocite{balestra2007} from a flat distribution, given a certain scatter in Fe abundance and systematic uncertainties? Suppose that we assume that the systematic uncertainties do not strongly depend on redshift and that the systematic uncertainty and scatter average out if we look at many different clusters. If we choose an exposure time for each cluster that will give a statistical error that is of the same order as the systematic one, then we can estimate how many clusters we need (assuming that the uncertainty on the average scales as $\sigma/\sqrt{N}$) and subsequently calculate the total needed exposure time. We perform this test by generating random abundance measurements following the power-law relation derived by Balestra et al. (2007)\nocite{balestra2007}: $Z_\text{Fe}(z) \simeq 0.54 (1+z)^{-1.25}$. We fit the simulated data points with a flat distribution and calculate the P-value for the $\chi^2$ value and the degrees of freedom. If the fit is acceptable, we count it as a detection of a flat distribution. Using this Monte Carlo approach, we determine the minimal amount of clusters needed to obtain a 90\% probability of finding an unacceptable fit, which translates in a 90\% chance of detecting evolution. If we assume an optimistic systematic scatter of 0.15 solar on the iron abundance, we would need observations of about 150 clusters in the 0.3 $< z <$ 1.0 range with a total exposure time of $\sim$13.7 Ms.     

Finally, a key question is whether such a study can be extended to other elements as well. In particular, the Si abundance has the advantage to be well constrained in the ICM (e.g. Rasia et al. 2008; Mernier et al. 2016a)\nocite{rasia2008,mernier2016a} and may be used to probe SNcc products over cosmic time. Unfortunately, our EPIC simulations show that even for 1 Ms of individual net exposure, Si cannot be constrained with less than $\sim 20 \%$ of accuracy for clusters with $z \gtrsim 0.5$. This illustrates the current limitations of CCD spectrometers, and the crucial importance of relying on micro-calorimeter technology for upcoming X-ray missions (Sections \ref{sec:Astro-H2} and \ref{sec:Athena}).

\subsection{\textit{Astro-H2}}
\label{sec:Astro-H2}

The next generation of high-resolution soft X-ray spectrometers is non-dispersive and very suitable for extended sources. The SXS instrument aboard the Japanese \textit{Hitomi} mission (\textit{Astro-H}, Takahashi et al., 2014)\nocite{takahashi2014} had a very high spectral resolution of $\sim$5 eV, a field of view of 3x3 arcmin, and a PSF of $\sim$1.2 arcmin, which allowed for high-resolution spectroscopy with limited spatial information. During its, unfortunately, short lifetime, SXS recorded an amazingly detailed spectrum of the Perseus cluster in the 2$-$10 keV band (Hitomi collaboration, 2016)\nocite{hitomi2016}. This spectrum shows the great potential of this type of spectrometer for bright and nearby clusters. An overview of the capabilities of \textit{Astro-H} in the cluster field is given in Kitayama et al. (2014)\nocite{kitayama2014}. The only limitations of \textit{Astro-H} are the limited PSF and effective area, which prevents the study of high-redshift clusters. Because \textit{Hitomi} showed that it can reveal new spectroscopic information about bright local clusters in unprecedented detail, a re-flight of the satellite (\textit{Astro-H2}) is expected to produce excellent spectra for cluster science.

\subsection{\textit{ATHENA}}
\label{sec:Athena}

At the end of the 2020s, ESA plans to launch its next generation X-ray observatory \textit{ATHENA} with the X-IFU spectrometer on board (Barret et al., 2013)\nocite{barret2013}. The effective area and PSF will be significantly better for \textit{ATHENA} then for \textit{Astro-H2}, which allows a deep study of high-redshift clusters in the $z=0.5-2$ range. Pointecouteau et al. (2013)\nocite{pointecouteau2013} show that the X-IFU instrument will be able to measure the oxygen, silicon, and iron abundance at least up to $z=1$ in 100 ks. For higher redshifts, a longer exposure is needed to detect oxygen, depending on the sensitivity of the mirrors at low energies. For local clusters, \textit{ATHENA} will be interesting to measure rare elements, like Na, Al, P, and Ti. The more element abundances can be measured, the better the constraints on the supernova models will be. Ettori et al. (2013)\nocite{ettori2013} show in more detail what \textit{ATHENA} will do for the physics of local clusters.

\section{Conclusions}

Given the systematic uncertainties in the abundance determination, which can be substantial (20$-$30\%), it is not useful to add a lot more exposure time to the current samples of local clusters of galaxies. Nevertheless, long exposures can be helpful in estimating and sometimes solving systematic errors. However, the current archive already contains a sufficient amount of long observations to reach the systematic limit. We have investigated the possibility to extend the cluster samples to higher redshift. With a substantial investment in exposure time of $\sim$14 Ms with \textit{XMM-Newton} it may be possible to differentiate between a possible power-law shaped evolution in Fe with redshift and a flat distribution. This is assuming that the systematic uncertainties average out when enough objects are observed. It will be hard and maybe not even possible to find a significant evolution in Fe with redshift with \textit{XMM-Newton}. Future missions like \textit{ATHENA} which are more sensitive and have a higher spectral resolution will be much more suited to study abundance evolution as a function of redshift.

\acknowledgements
The Netherlands Institute for Space Research (SRON) is supported financially
by NWO, the Netherlands Organization for Scientific Research.

\newpage%%%%%%%%%%%%%%%%%%%%%%%%%%%%%%%%%%%%%%%%%%%%%%%%%%%%%%

\bibliographystyle{an}
\bibliography{jdeplaa}

\begin{thebibliography}{}

\bibitem{ahoranta2016}
{Ahoranta}, J., {Finoguenov}, A., {Pinto}, C., {Sanders}, J., {Kaastra}, J.,
  {de Plaa}, J., {Fabian}, A.: 2016,  ArXiv: 1607.07444

\bibitem{anderson2009}
{Anderson}, M.\,E., {Bregman}, J.\,N., {Butler}, S.\,C., {Mullis}, C.\,R.:
  2009,  \apj 698, 317--323

\bibitem{badenes2006}
{Badenes}, C., {Borkowski}, K.\,J., {Hughes}, J.\,P., {Hwang}, U., {Bravo}, E.:
  2006,  \apj 645, 1373--1391

\bibitem{baldi2012}
{Baldi}, A., {Ettori}, S., {Molendi}, S., {Balestra}, I., {Gastaldello}, F.,
  {Tozzi}, P.: 2012,  A\&A 537, A142

\bibitem{balestra2007}
{Balestra}, I., {Tozzi}, P., {Ettori}, S., {Rosati}, P., {Borgani}, S.,
  {Mainieri}, V., {Norman}, C., {Viola}, M.: 2007,  A\&A 462, 429--442

\bibitem{barret2013}
{Barret}, D., {den Herder}, J.\,W., {Piro}, L., et~al.: 2013,  ArXiv: 1308.6784

\bibitem{boehringer2010}
{B{\"o}hringer}, H., {Werner}, N.: 2010,  A\&ARv 18, 127--196

\bibitem{bravo2012}
{Bravo}, E., {Mart{\'{\i}}nez-Pinedo}, G.: 2012,  Phys. Rev. C 85(5), 055805

\bibitem{bromm2004}
{Bromm}, V., {Larson}, R.\,B.: 2004,  \araa 42, 79

\bibitem{deplaa2013}
{de Plaa}, J.: 2013,  Astronomische Nachrichten 334, 416

\bibitem{deplaa2016}
{de Plaa}, J.: 2016,  A\&A {subm.}

\bibitem{deplaa2007}
{de Plaa}, J., {Werner}, N., {Bleeker}, J.\,A.\,M., {Vink}, J., {Kaastra},
  J.\,S., {M{\'e}ndez}, M.: 2007,  A\&A 465, 345--355

\bibitem{ettori2013}
{Ettori}, S., {Pratt}, G.\,W., {de Plaa}, J., et~al.: 2013,  ArXiv: 1306.2322

\bibitem{hitomi2016}
{Hitomi Collaboration}, {Aharonian}, F., {Akamatsu}, H., {Akimoto}, F., {et
  al.}: 2016,  Nature 535, 117

\bibitem{iwamoto1999}
{Iwamoto}, K., {Brachwitz}, F., {Nomoto}, K., {Kishimoto}, N., {Umeda}, H.,
  {Hix}, W.\,R., {Thielemann}, F.: 1999,  \apjs 125, 439

\bibitem{kaastra1996}
{Kaastra}, J.\,S., {Mewe}, R., {Nieuwenhuijzen}, H.: 1996, {SPEX: a new code
  for spectral analysis of X \& UV spectra.}
\newblock In {Yamashita}, K., {Watanabe}, T. (Eds.), UV and X-ray Spectroscopy
  of Astrophysical and Laboratory Plasmas (pp.\, 411--414)

\bibitem{kitayama2014}
{Kitayama}, T., {Bautz}, M., {Markevitch}, M., et~al., {on behalf of the
  ASTRO-H Science Working Group}: 2014,  ArXiv: 1412.1176

\bibitem{madau2014}
{Madau}, P., {Dickinson}, M.: 2014,  \araa 52, 415--486

\bibitem{matteucci2005}
{Matteucci}, F., {Calura}, F.: 2005,  \mnras 360, 447

\bibitem{maughan2008}
{Maughan}, B.\,J., {Jones}, C., {Forman}, W., {Van Speybroeck}, L.: 2008,
  \apjs 174, 117--135

\bibitem{mernier2016a}
{Mernier}, F., {de Plaa}, J., {Pinto}, C., {Kaastra}, J.\,S., {Kosec}, P.,
  {Zhang}, Y.\,Y., {Mao}, J., {Werner}, N.: 2016a,  A\&A 592, A157

\bibitem{mernier2016b}
{Mernier}, F., {de Plaa}, J., {Pinto}, C., {Kaastra}, J.\,S., {Kosec}, P.,
  {Zhang}, Y.\,Y., {Mao}, J., {Werner}, N., {Pols}, O.\,R., {Vink}, J.: 2016b,
  A\&A in press, ArXiv: 1608.03888

\bibitem{mushotzky1997}
{Mushotzky}, R.\,F., {Loewenstein}, M.: 1997,  \apjl 481, L63--L66

\bibitem{nomoto2006}
{Nomoto}, K., {Tominaga}, N., {Umeda}, H., {Kobayashi}, C., {Maeda}, K.: 2006,
  Nuclear Physics A 777, 424

\bibitem{pinto2014}
{Pinto}, C., {Fabian}, A.\,C., {Werner}, N., {Kosec}, P., {Ahoranta}, J., {de
  Plaa}, J., {Kaastra}, J.\,S., {Sanders}, J.\,S., {Zhang}, Y.\,Y.,
  {Finoguenov}, A.: 2014,  A\&A 572, L8

\bibitem{pinto2015}
{Pinto}, C., {Sanders}, J.\,S., {Werner}, N., {de Plaa}, J., {Fabian}, A.\,C.,
  {Zhang}, Y.\,Y., {Kaastra}, J.\,S., {Finoguenov}, A., {Ahoranta}, J.: 2015,
  A\&A 575, A38

\bibitem{pointecouteau2013}
{Pointecouteau}, E., {Reiprich}, T.\,H., {Adami}, C., et~al.: 2013,  ArXiv:
  1306.2319

\bibitem{rasia2008}
{Rasia}, E., {Mazzotta}, P., {Bourdin}, H., {Borgani}, S., {Tornatore}, L.,
  {Ettori}, S., {Dolag}, K., {Moscardini}, L.: 2008,  \apj 674, 728--741

\bibitem{schellenberger2015}
{Schellenberger}, G., {Reiprich}, T.\,H., {Lovisari}, L., {Nevalainen}, J.,
  {David}, L.: 2015,  A\&A 575, A30

\bibitem{takahashi2014}
{Takahashi}, T., {Mitsuda}, K., {Kelley}, R., et~al.: 2014, {The ASTRO-H X-ray
  astronomy satellite}.
\newblock In Space Telescopes and Instrumentation 2014: Ultraviolet to Gamma
  Ray, Volume 9144 of SPIE Proc. (p.\, 914425)

\bibitem{tozzi2003}
{Tozzi}, P., {Rosati}, P., {Ettori}, S., {Borgani}, S., {Mainieri}, V.,
  {Norman}, C.: 2003,  \apj 593, 705--720

\bibitem{waldman2011}
{Waldman}, R., {Sauer}, D., {Livne}, E., {Perets}, H., {Glasner}, A.,
  {Mazzali}, P., {Truran}, J.\,W., {Gal-Yam}, A.: 2011,  \apj 738, 21

\bibitem{werner2008}
{Werner}, N., {Durret}, F., {Ohashi}, T., {Schindler}, S., {Wiersma},
  R.\,P.\,C.: 2008,  SSRv 134, 337--362

\bibitem{willingale2013}
{Willingale}, R., {Starling}, R.\,L.\,C., {Beardmore}, A.\,P., {Tanvir},
  N.\,R., {O'Brien}, P.\,T.: 2013,  MNRAS 431, 394--404

\bibitem{woosley1995}
{Woosley}, S.\,E., {Weaver}, T.\,A.: 1995,  \apjs 101, 181

\end{thebibliography}

\end{document}